\documentclass[12pt]{article}
\usepackage{hyperref}
\usepackage{amssymb,amsmath,graphicx,latexsym}%\pdfoutput=1 %jai supprimé 'cite' dans le package

\begin{document}
\title{Modified gravity from a functional of entropy}
\author{Fay\c{c}al Hammad\thanks{fayhammad@gmail.com}\\
\emph{\small D\'{e}partment ST, Universit\'{e} A.Mira,}\\[-1ex]
\emph{\small 06000 Bejaia, Algeria}}
\date{}
\maketitle
\begin{abstract}
We extend Padmanabhan's entropy functional formalism to show that, in addition to the Gauss-Bonnet or the entire series of Lanczos-Lovelock Lagrangians already obtained, more general higher-order corrections to General Relativity, i.e., the so-called modified gravity theories, also emerge naturally from this formalism. This extension shows that the formalism constitutes a valuable tool to investigate, at each order in the curvature, the possible structure the higher-order modified gravity theories might have. As an application, the extended formalism is used to evaluate the horizon entropy in a modified gravity theory of the second-order in the curvature. Our findings are in agreement with previous results from the literature.
\end{abstract}

\begin{quote}
PACS numbers: 04.50.-h, 04.70.Dy, 83.10.Ff.
\newline
Keywords: {\em Modified gravity, continuum mechanics, black hole thermodynamics.}
\end{quote}


%\noindent{\it Keywords}: Elasticity, entropy, cosmic expansion
%\pacs{46.25.-y, 05.70.-a, 98.80.Cq}

\section{Introduction}\label{sec:1}
Recently, it became increasingly clear that there is a real need for a modified theory of gravity, and much insight regarding the possible ways to modify General Relativity, as well as the advantages of doing so, has been gained (see e.g. the reviews \cite{Reviews}). The simplest modification is the so-called scalar-tensor or $\mathcal{F}(\phi,R)$-modified gravity theories \cite{Faraoni,CapoFar}. In this class of theories, no scalar other than those formed from the Ricci scalar $R$ and an independent scalar field $\phi$ intervenes \footnote{see e.g. \cite{HammadVPL} for a more recent model belonging to this class of modified gravity and its applications for cosmology.}. However, fundamental approaches like string theory \cite{GSW}, or the study of curved spacetime quantum field theory \cite{BirellDavis}, had already imposed specific higher-order corrections to General Relativity. The most familiar correction, second-order in the curvature, being the Gauss-Bonnet (GB) topological invariant $\mathcal{R}^{2}_{GB}$ coupled to a dilaton/modulus field \cite{GBFromStrings}. It was extensively used in the study of black holes \cite{GBBH}, as well as the early expansion of the Universe \cite{GBInflation} and its late time expansion \cite{GBDE}.

Nevertheless, it turned out that constructing a general scalar-Gauss-Bonnet Lagrangian might also explain the actually observed features of our Universe, such as dark energy \cite{SGBDE}, as well as to find alternative origins to the early inflation \cite{SGBInflation}. In addition, an interesting combination of a functional of the Ricci scalar with a scalar-Gauss-Bonnet term is investigated in \cite{f(R)+SGB} and interesting consequences of a general functional of the GB invariant for cosmology can be found in \cite{f(GB)}. What's more, some authors have also shown the possibility, and the advantages for cosmology, of adding to the Einstein-Hilbert action a functional of the three quadratic terms contained in the GB invariant, each taken with a different weight \cite{R2+P+Q}.

The study of the thermodynamics of spacetime, on the other hand, turned out to be a valuable tool for the investigation of its dynamics as well. Indeed, it turned out that one is able to use thermodynamics to re-derive General Relativity itself \cite{Jacobson}. In fact, it was even suggested that Newton's laws \cite{Verlinde} as well as the cosmic expansion \cite{EassFramSmoo} might be explained out using entropic arguments.

In fact, by constructing an entropy functional, T. Padmanabhan has also been able to re-derive the field equations of General Relativity \cite{Pad}. In \cite{HammadCQG}, it was found that when the latter formalism is coupled with a generalized four-dimensional elasticity theory, both the early and the late time expansions of the Universe might be accounted for. In Padmanabhan's formalism, one takes the macroscopic deformations of spacetime to be a manifestation of its microscopic 'structure' and introduces a vector field to represent these deformations. One then associates to this deformation, or displacement, vector field, in analogy with elasticity theory of classical three-dimensional media, an entropy functional. In \cite{Pad+Par,Pad2}, it was shown that demanding that a given precise structure of the entropy functional be extremal in this vector field is sufficient to recover the field equations of the entire series of Lanczos-Lovelock Lagrangians \cite{Lanczos}, in which General Relativity appears at the first order and the GB Lagrangian appears at the second order.

In \cite{Hammad+Erratum}, an extended entropy functional formalism was introduced to include spacetimes with torsion. It was shown there that demanding that the extended functional be extremal in the displacement vector field is sufficient to recover the Cartan-Sciama-Kibble field equations of Einstein-Cartan gravity.

In the present paper, we extend further the latter formalism and show that more general higher-order modified gravity theories (we shall not include torsion in this paper, however) are recovered when demanding that the functional be extremal. The formalism imposes specific corrections to be brought to the Einstein-Hilbert action at each order in the curvature, narrowing thereby the range of possibilities one might think of.

This paper is organized as follows. In Sec.~\ref{sec:2}, we recall, for later reference, separately the field equations of $\mathcal{F}(\phi,R)$-modified gravity as well as of modified gravity with correction terms quadratic in the curvature. In Sec.~\ref{sec:3}, we show how $\mathcal{F}(\phi,R)$-modified gravity emerges from a simple extension of the entropy functional formalism. In Sec.~\ref{sec:4}, we argue how consistency in the formalism does not allow to have only $\mathcal{F}(\phi,R)$ theories not containing other curvature invariants built from the Ricci and the Riemann tensors. We then motivate and construct the entropy functional at the zeroth order in the Newton's constant $G_{N}$, from which the field equations of a modified gravity of a second-order in the curvature emerge. The gravitational Lagrangian of the corresponding modified gravity is found. An outline of the procedure for dealing with higher-order corrections is given in Sec.~\ref{sec:5}. The calculation of a black hole entropy using this extended formalism is exposed in Sec.~\ref{sec:6}. Section \ref{sec:7} is devoted to a detailed discussion of the differences and the similarities our extended formalism has compared with the original entropy functional formalism of Padmanabhan et al. and to highlight new subtle features pertaining to the former. We end this paper with a brief summary and discussion section.

\section{Some modified gravity formalism}\label{sec:2}
In this section we recall, for later reference, the actions and the field equations of well-known modified gravity theories that we shall recover in the subsequent sections. We will be using, throughout the paper, units where $\hbar=c=k_{B}=1$, where $k_{B}$ is the Boltzmann constant.

We begin with the scalar-tensor gravity theories in which the gravitational sector is given by a functional $\mathcal{F}(\phi,R)$, where $\mathcal{F}$ is a regular but otherwise arbitrary functional of the Ricci scalar $R$ and an independent scalar field $\phi$ \cite{Faraoni,CapoFar}:
\begin{equation}\label{1}
S=\int\mathrm{d}^{4}x\sqrt{-g}\left[\mathcal{F}(\phi,R)+\mathcal{L}_{\mathrm{matter}}\right],
\end{equation}
where $\mathcal{L}_{\mathrm{matter}}=\mathcal{L}(g_{\mu\nu},\psi)$ is the Lagrangian of the matter fields $\psi$. The equations of motion when varying the action with respect to the scalar field and the metric are, respectively,
\begin{align}\label{2}
\frac{\delta\mathcal{F}}{\delta\phi}&=0,\nonumber
\\T_{\mu\nu}&=\frac{\partial\mathcal{F}}{\partial R}R_{\mu\nu}-\frac{1}{2}g_{\mu\nu}\mathcal{F}-\nabla_{\mu}\nabla_{\nu}\frac{\partial \mathcal{F}}{\partial R}+g_{\mu\nu}\Box\frac{\partial \mathcal{F}}{\partial R},
\end{align}
where $\Box$ is the d'Alambertian covariant operator. Note that here we assume, as is usually done in the literature, that the field $\phi$ does not couple directly to matter, but only through its non-minimal coupling with geometry.

Another well-known family of modified gravity theories is the Gauss-Bonnet fourth-order gravity \cite{CapoFar}. In this family, one uses the Gauss-Bonnet invariant $\mathcal{R}^{2}_{GB}=R^{2}-4R_{\mu\nu}R^{\mu\nu}+R_{\mu\nu\rho\sigma}R^{\mu\nu\rho\sigma}$. At this order in the curvature, however, this particular combination of the quadratic invariants does not exhaust all the possible scalars one might construct. Indeed, it is possible to have arbitrary algebraic combinations of the previous three quadratic invariants inside the GB term, each weighted with a different spacetime-dependent factor, and not excluding also terms of the form $\partial^{\mu}\phi\partial_{\mu}R$ and $\Box R$. Therefore, a general action principle one might write down at the second order in the curvature is of the form
\begin{equation}\label{3}
S=\int\mathrm{d}^{4}x\sqrt{-g}\left[\mathcal{F}(\phi,R,P,Q)+\mathcal{L}_{\mathrm{matter}}\right],
\end{equation}
where $\mathcal{F}$ is a regular, but arbitrary, functional of the scalars $\phi$, $R$, $P\equiv R_{\mu\nu}R^{\mu\nu}$, and $Q\equiv R_{\mu\nu\rho\sigma}R^{\mu\nu\rho\sigma}$. Note that elsewhere nonlocal terms such as $R\Box^{-1}R$ are not excluded \cite{Reviews}. Here, however, we restrict ourselves to local terms but we will come back to this issue later. The equations of motion one obtains from this action are
\begin{align}\label{4}
\frac{\delta\mathcal{F}}{\delta\phi}&=0,\nonumber
\\T_{\mu\nu}&=g_{\mu\nu}\left[-\frac{1}{2}\mathcal{F}+\Box\mathcal{F}_{R}+\nabla_{\rho}\nabla_{\sigma}\left(R^{\rho\sigma}\mathcal{F}_{P}\right)\right]
+R_{\mu\nu}\mathcal{F}_{R}+2R_{\mu\rho}{R^{\rho}}_{\mu}\mathcal{F}_{P}-\nabla_{\mu}\nabla_{\nu}\mathcal{F}_{R}\nonumber
\\&+2R_{\mu\rho\sigma\tau}{R_{\nu}}^{\rho\sigma\tau}\mathcal{F}_{Q}
-2\nabla_{\rho}\nabla_{(\mu}\left[{R^{\rho}}_{\nu)}\mathcal{F}_{P}\right]+\Box\left(R_{\mu\nu}\mathcal{F}_{P}\right)
-4\nabla^{\rho}\nabla^{\sigma}\left[R_{\mu(\rho\sigma)\nu}\mathcal{F}_{Q}\right],
\end{align}
where, as it will be the convention throughout this paper, indices inside round brackets mean a symmetrization with 'weight 1' in those indices, whereas indices inside square brackets mean an antisymmetrization with 'weight 1' in those indices. Also, for notational convenience, we have denoted, as is customary, $\mathcal{F}_{R}\equiv\partial\mathcal{F}/\partial R$, $\mathcal{F}_{P}\equiv\partial\mathcal{F}/\partial P$ and $\mathcal{F}_{Q}\equiv\partial\mathcal{F}/\partial Q$.

\section{$\mathcal{F}(\phi,R)$-modified gravity}\label{sec:3}
The basic idea behind the entropy functional formalism \cite{Pad}, as recalled in the introduction, is to associate to spacetime a displacement vector field $u^{\mu}(x)$ such that $\bar{v}^{\mu}={v}^{\mu}+u^{\mu}(x)$, where $\bar{v}^{\mu}$ and $v^{\mu}$ are coordinate labels of spacetime events after and before the deformation, respectively, and then construct the corresponding entropy functional. The functional should be a scalar quadratic in the field $u^{\mu}$ as well as its first derivatives, in order not to get more than second-order differential equations of motion for the field $u^{\mu}$. Furthermore, in order to obtain linear differential equations of motion, the functional must not contain more than quadratic terms in the field. The precise form of each term inside the functional is then dictated by an analogy with elasticity theory of three-dimensional media. Matter in this formalism is viewed as a defect that spoils translational invariance in the field $u^{\mu}$, a fact that translates inside the functional into a coupling of the energy-momentum tensor of matter with two components of the field $u^{\mu}$. To this term, one may add \cite{HammadBH} another possible term proportional to $u_{\mu}u^{\mu}$ that can be interpreted as a background-dependent 'potential energy' of the field $u^{\mu}$. For the terms quadratic in the derivatives, one allows for every possible contraction of two derivatives $\nabla^{\mu}u^{\nu}$ among themselves. Based on these arguments, the functional started with in \cite{HammadBH} had the following form,
\begin{align}\label{5}
{\cal S}=\int_{\mathcal{V}}\mathrm{d}^{4}x\sqrt{-g}\Big[&A\nabla_{\mu}u_{\nu}\nabla^{\nu}u^{\mu}+B\nabla_{\mu}u_{\nu}\nabla^{\mu}u^{\nu}
+C\left(\nabla_{\mu}u^{\mu}\right)^{2}+\left(\lambda g_{\mu\nu}+T_{\mu\nu}\right)u^{\mu}u^{\nu}\Big],
\end{align}
where $\mathcal{V}$ is the spacetime region under study. $A$, $B$, and $C$ were three constants, and $\lambda$ an arbitrary spacetime-dependent scalar. It was found in \cite{HammadBH} that in order for this functional to be extremal for every displacement field $u^{\mu}$, one must impose the constraints, $A=-C$ and $B=0$. Then $A$ was chosen to be $1/8\pi G_{N}$, where $G_{N}$ is Newton's constant, in order to recover the newtonian limit of General Relativity. $\lambda$ was found to be given by $2\lambda=AR-2\Lambda$ where $R$ is the Ricci scalar and $\Lambda$ an integration constant, interpreted, in accordance with Ref.~\cite{Pad}, as a cosmological constant.

Now, as a first generalization of this functional, we simply relax the assumption made from the outset that the three factors $A$, $B$ and $C$ are all constant. Namely, we just let these be, more naturally, spacetime-dependent, that is, scalar fields. An analogy with three-dimensional elasticity theory would be to allow for position- and time-dependent elastic constants \cite{WWGY} and kinetic components.

The second generalization will be, as in \cite{Hammad+Erratum}, the allowance to have terms of the form $u^{\mu}\nabla^{\nu}u^{\rho}$ in the functional, but exclude higher products of the vector $u^{\mu}$ and higher-order derivatives in order for the functional to yield at most second-order \textit{linear} differential equations of motion for the field $u^{\mu}$. In \cite{Hammad+Erratum}, these additional terms were motivated by the possibility of coupling the field $u^{\mu}$ with spin-angular momentum tensor $\Sigma_{\mu\nu\rho}$ of matter with intrinsic spin, since the main assumption of the whole entropy functional approach consists in viewing matter as defaults within the spacetime continuum that breaks translational invariance in the displacement vector field $u^{\mu}$. Here, we introduce the $u^{\mu}\nabla^{\nu}u^{\rho}$ terms, even in the absence of matter with intrinsic spin, by making the natural assumption that translational invariance may also be broken whenever a scalar field with a non-vanishing gradient arises within the spacetime continuum. To make an analogy with continuum mechanics, it is the appearance of vortices somewhere inside a fluid whenever a gradient of a flow is present (see e.g. \cite{Landau}). Therefore, this analogy also suggests that even the gradients of scalar fields pertaining to the medium itself might contribute to this coupling. In the case of spacetime, the Ricci scalar is the simplest scalar that could be built from the background geometry. Another possibility is to have an additional scalar field, usually called $\phi$, that pertains to the geometry of spacetime but is independent of its metric or curvature.

Yet, another possible interpretation of these additional terms would be the allowance of 'friction' terms to contribute inside the functional. Indeed, when varying the functional with respect to vector field $u^{\mu}$, one formally obtains, thanks to these additional terms, equations of motion of the form $a\partial^{2}u+b\partial u+cu=0$. The coefficient $b$ is interpreted in classical mechanics as being responsible for the existence of friction. In fluid mechanics, it gives rise to the fluid's viscosity and may also be due to thermal conduction \cite{Landau}. In our case, that coefficient comes from the presence of scalar field gradients, whence their suggested friction interpretation.

All this amounts then to assume that in this generalized functional the term $u^{\mu}\nabla^{\nu}u^{\rho}$ should also couple with the gradient of scalar fields built from the geometry of the medium such as the Ricci scalar $R$, and/or an independent scalar field $\phi$. Hence, the resulting general form of the extended entropy functional will be
\begin{align}\label{6}
{\cal S}=\int_{\mathcal{V}}\mathrm{d}^{4}x\sqrt{-g}\Big[&\left(Ag_{\mu\sigma}g_{\nu\rho}+Bg_{\mu\rho}g_{\nu\sigma}
+Cg_{\mu\nu}g_{\rho\sigma}\right)\nabla^{\mu}u^{\nu}\nabla^{\rho}u^{\sigma}\nonumber
\\&+\left(D_{,\mu}g_{\nu\rho}+E_{,\nu}g_{\mu\rho}+F_{,\rho}g_{\mu\nu}\right)u^{\mu}\nabla^{\nu}u^{\rho}+\left(\lambda g_{\mu\nu}+T_{\mu\nu}\right)u^{\mu}u^{\nu}\Big],
\end{align}
where $A$,..., $F$ are all scalar functionals of the Ricci scalar $R$ and/or an additional scalar field $\phi$, and their derivatives. For notational convenience, a comma in front of a letter will denote throughout the paper a covariant derivative. Also, in this paper we shall restrict ourselves to a single independent scalar field $\phi$ but the approach may readily be generalized to include multiplets of scalar fields.

Now, we would like to make the following important remark before we proceed further. By dimensional analysis, we know that if we take the vector field $u^{\mu}$, as well as the scalar field $\phi$, to be both dimensionless, in order for the entropy functional to be dimensionless too, we must divide or multiply each of the above terms by the necessary powers of Newton's constant $G_{N}$ so as to cancel the dimensions brought by each of the above scalar functionals. For instance, if one chooses the scalar $A$ to be constant, as it was done in \cite{HammadBH}, one needs to have a factor proportional to $1/G_{N}$ in front of the first term $\nabla^{\mu}u^{\nu}\nabla^{\rho}u^{\sigma}$, since the latter has the dimensions of $(\mathrm{length})^{-2}$, in order to cancel the dimension $(\mathrm{length})^{4}$ of the volume element $\mathrm{d}^{4}x\sqrt{-g}$. If, on the other hand, one allows to have positive powers of the Ricci scalar and/or the scalar field $\phi$ and their derivatives inside the scalar functionals, one would have inside the entropy integral terms proportional to positive powers of $G_{N}$. Hence, the precise form of these scalar functionals one chooses actually determines the order of approximation in powers of $G_{N}$ one wishes to achieved inside the entropy integral. Accordingly, as we shall see below, fixing the approximation level inside the entropy functional determines precisely the equations of motion one obtains for the spacetime background and, thereby, the required corrections to bring to the Einstein-Hilbert action at that level.

Varying the functional (\ref{6}) with respect to the field $u^{\mu}$, with vanishing variations on the boundaries, and then integrating by parts gives
\begin{align}\label{7}
\delta{\cal S}=\int_{\mathcal{V}}\mathrm{d}^{4}x\sqrt{-g}\Big[&-\big(2A_{,\sigma}g_{\nu\rho}+2B_{,\rho}g_{\nu\sigma}+2C_{,\nu}g_{\rho\sigma}
-D_{,\nu}g_{\rho\sigma}-F_{,\sigma}g_{\nu\rho}+D_{,\sigma}g_{\rho\nu}+F_{,\nu}g_{\rho\sigma}\big)\nabla^{\rho}u^{\sigma}\nonumber
\\&-2\left(A\nabla_{\mu}\nabla_{\nu}u^{\mu}+B\Box u_{\nu}+C\nabla_{\nu}\nabla_{\mu}u^{\mu}\right)\nonumber
\\&+\left(2\lambda g_{\mu\nu}+2T_{\mu\nu}-D_{,\mu\nu}-g_{\mu\nu}\Box E-F_{,\mu\nu}\right)u^{\mu}\Big]\delta u^{\nu}.
\end{align}
Then, the condition $\delta{\cal S}=0$ for all variations $\delta u^{\mu}$ implies the vanishing of everything that is inside the square brackets of (\ref{7}). This fact, in turn, becomes possible without having to impose any constraint on the field $u^{\mu}$ if, \textit{a priori}, the content of each of the parentheses inside the square brackets vanishes separately. The vanishing of the first parenthesis gives
\begin{equation}\label{8}
g_{\nu\rho}(2A+D-F)_{,\sigma}+g_{\rho\sigma}(2C+F-D)_{,\nu}+2g_{\nu\sigma}B_{,\rho}=0,
\end{equation}
which is identically satisfied if $2A+D-F=\mathrm{const}$, $2C+F-D=\mathrm{const}$ and $2B=\mathrm{const}$. This result actually turns the initial constraint of having all three parentheses inside integral (\ref{7}) vanish separately into a weaker condition. Indeed, the first and the second term of the second parenthesis of (\ref{7}) may yield a single linear term in the vector field $u^{\mu}$ provided that $C=-A$, for then one might use the identity, $2\nabla_{[\nu}\nabla_{\mu]}u^{\nu}=R_{\mu\nu}u^{\nu}$, to get rid of the derivatives. This allows the content of the third parenthesis to be fused with that of the second one just by adding the term $-2AR_{\mu\nu}$ to the former. Thus, the necessary and sufficient condition is to have, besides $2A+D-F=\mathrm{const}$, the following single constraint
\begin{equation}\label{9}
2\lambda g_{\mu\nu}+2T_{\mu\nu}-D_{,\mu\nu}-g_{\mu\nu}\Box E-F_{,\mu\nu}-2AR_{\mu\nu}=0.
\end{equation}
Finally, the remaining term $B\Box u_{\nu}$ in the second parenthesis implies that $B$ is actually a vanishing constant. Next, substituting for $D$ in (\ref{9}) the value $F-2A+\mathrm{const}$, deduced from the constraint obtained above, yields
\begin{equation}\label{10}
2T_{\mu\nu}=2AR_{\mu\nu}-2A_{,\mu\nu}+2F_{,\mu\nu}+g_{\mu\nu}\Box E-2\lambda g_{\mu\nu}.
\end{equation}

Taking the four-divergence of the latter equation and using the conservation equation $\nabla^{\mu}T_{\mu\nu}=0$ of the energy-momentum tensor of matter, we obtain
\begin{equation}\label{11}
0=2{R^{\mu}}_{\nu}F_{,\mu}+\left(\Box[2F-2A+E]-2\lambda\right)_{,\nu}+AR_{,\nu},
\end{equation}
where we have used $\nabla^{\mu}R_{\mu\nu}=\frac{1}{2}\nabla_{\nu}R$, as well as the fact that $\Box\nabla_{\mu}f=R_{\mu\nu}\nabla^{\mu}f+\nabla_{\nu}\Box f$ for every scalar $f$, in order to transform the resulting third-order differential equation into a combination of gradients. Since the above equation, in which intervene only geometric quantities inside the scalars $A$, $E$, $F$, and $\lambda$, must be satisfied without introducing any constraints other than the Bianchi identities, each category of terms must vanish separately. The factor multiplying the Ricci tensor should vanish identically, implying that $F=\mathrm{const}$. But since only the gradient of the scalar $F$ appears inside the entropy integral, this condition is equivalent to simply having $F=0$. The vanishing of the remainder of the right-hand side of (\ref{11}) then yields
\begin{align}\label{12}
\frac{\delta}{\delta\phi}\left(2\Box A-\Box E+2\lambda\right)&=0,\nonumber
\\\frac{\partial}{\partial R}\left(2\Box A-\Box E+2\lambda\right)&=A.
\end{align}
Therefore, due to the extremality requirement $\delta\mathcal{S}=0$ for every vector $u^{\mu}$, not all the scalar functionals we started with are independent, and some of them need not even appear inside the entropy functional.

From the remark made below integral (\ref{6}), it follows that it is at this point that a given $\mathcal{F}(\phi,R)$-modified gravity theory will result from the formalism after one chooses the approximation level one wishes to achieve inside the entropy integral.

Indeed, restraining oneself for example to the approximation level $G^{-1}_{N}$, the scalar functionals $A$ and $E$ may only depend on $\phi$, while the scalar $\lambda$, since it appears multiplied simply by the field $u^{\mu}$ but not its derivatives, may depend on $\phi$, its second derivative, and contain one positive power of $R$. Therefore, the second partial differential equation in (\ref{12}) integrates to $2\Box A-\Box E+2\lambda=A(\phi)R-2\Lambda(\phi)$, for some functional $\Lambda(\phi)$. The first equation of (\ref{12}) then implies that $R\delta A(\phi)/\delta\phi=2\delta\Lambda(\phi)/\delta\phi$. Substituting these results into $(\ref{10})$, the latter simply reads
\begin{align}\label{13}
T_{\mu\nu}=A(\phi)\left(R_{\mu\nu}-\frac{1}{2}g_{\mu\nu}R\right)+g_{\mu\nu}\Lambda(\phi)-\nabla_{\mu}\nabla_{\nu}A(\phi)+g_{\mu\nu}\Box A(\phi).
\end{align}
This represents the usual set of Einstein's field equations with a scalar field-dependent gravitational constant and a cosmological constant $\Lambda(\phi)$, in which the scalar field $\phi$ satisfies the equation of motion $R\delta A(\phi)/\delta\phi=2\delta\Lambda(\phi)/\delta\phi$. For the special case $A(\phi)=\phi$, one recovers the Brans-Dicke field equations, where $\Lambda(\phi)$ plays the role of the Brans-Dicke's scalar field potential \cite{Faraoni}, whereas the case $A(\phi)=\mathrm{const}$ gives back General Relativity with an ordinary scalar field $\phi$, provided that the constant $A$ is identified with $1/8\pi G_{N}$. Thus, at the order $G^{-1}_{N}$, one simply recovers either the usual General Relativity or the latter, non-minimally coupled to a scalar field $\phi$.

If, on the other hand, one allows to go up to the $G^{0}_{N}$-th approximation order inside (\ref{6}), one would have to allow the functionals $A$ and $E$ to depend not only on $\phi$, but also on $R$ and on the derivatives of $\phi$ up to the second order. As for $\lambda$, not being coupled to derivatives of the vector $u^{\mu}$, it may contain $\phi$, $R^{2}$, as well as their derivatives, up to the second order for $R$ and up to the fourth order for $\phi$. Hence, terms such as $\Box R$ and $\partial^{\mu}\phi\partial_{\mu}R$ may also appear. Note that this power counting does not exclude the appearance of the term $R\Box^{-1} R$, yielding what is called non-local modified gravity theories \cite{CapoFar}. We will elaborate more on to this remark in the last section.

More generally, we see by power counting that any chosen order in $G_{N}$ will impose a limit on the powers and the number of derivatives of $R$, as well as the number of derivatives of $\phi$, that one may include inside the scalars $A(\phi,R)$ and $E(\phi,R)$. Integrating again the second partial differential equation in (\ref{12}) with these general cases, we find $2\Box A-\Box E+2\lambda=\mathcal{F}(\phi,R)$, such that $A=\partial\mathcal{F}/\partial R$. The first equation of (\ref{12}) then implies that $\delta \mathcal{F}(\phi,R)/\delta\phi=0$. Substituting these results into $(\ref{10})$, we get exactly the complete set of field equations in (\ref{2}), i.e., $\mathcal{F}(\phi,R)$-modified gravity theories.

Before extending this formalism further in the next section, we would like to make the following remark that reveals a key element of this whole extension procedure of the entropy functional formalism. If we haven't allowed the scalar $\lambda$ to depend on a field $\phi$ and haven't introduced inside the entropy functional the additional hybrid terms $u\nabla u$, which is equivalent to having chosen the three scalars $D$, $E$, and $F$ all constant such that their gradients vanish, the above algebraic relations relating these scalars to $A$ would have given also a constant value for the latter, as well as a fixed integration constant $\Lambda$ for the second differential equation in (\ref{12}). It is then clear that in this extended formalism, functionals of the curvature scalar $R$ and the scalar $\phi$ arise as corrections to the Hilbert action thanks to the additional coupling of 'currents' inside the medium with the term $u\nabla u$; these 'currents' being the gradients of scalars belonging to the spacetime medium.

\section{More general second-order corrections}\label{sec:4}
In this section, we shall examine how a further generalization of the functional (\ref{6}) permits to obtain specific modified gravity theories of a second-order in the curvature. Up to now, we have allowed contractions of the gradients $u\nabla u$ only with the background metric tensor. It is however unnatural to exclude contractions with the Riemann or the Ricci tensors which are also built from the background geometry. Further, we have up to now allowed the scalar functionals inside the integral (\ref{6}) to be functionals only of the Ricci scalar $R$, and/or an independent scalar field $\phi$ and their derivatives. As it is well-known, however, one can build scalar fields also from contractions of the Ricci tensor $R_{\mu\nu}$ and the Riemann tensor $R_{\mu\nu\rho\sigma}$ among themselves. The most famous scalar built this way being the Gauss-Bonnet topological invariant $\mathcal{R}^{2}_{GB}$. Actually, in this formalism it would appear simply unnatural to include, at a given approximation in powers of $G_{N}$, higher powers of $R$ without also including products of the Riemann and the Ricci tensors. But, if one allows the contribution of scalars built from contractions of such tensors inside the entropy functional, one should also, for the sake of consistency, allow the contribution of those tensors un-contracted among themselves, as well as their gradients, as it follows from our discussion in the third paragraph below integral (\ref{5}). The procedure might then be much more involved than what has been done in the previous section. For this reason, we shall first examine in what follows the symmetries that the most general form of the entropy integral should possess and the constraints to be imposed on its components, in order to facilitate their detailed construction later.

Let us first write down the most general entropy integral that includes the 'friction' terms $u\nabla u$:
\begin{align}\label{14}
{\cal S}=\int_{\mathcal{V}}\mathrm{d}^{4}x\sqrt{-g}\Big(\Phi_{\mu\nu\rho\sigma}\nabla^{\mu}u^{\nu}\nabla^{\rho}u^{\sigma}
+\Psi_{\mu\nu\rho}u^{\mu}\nabla^{\nu}u^{\rho}+\Pi_{\mu\nu}u^{\mu}u^{\nu}\Big).
\end{align}
We note, first of all, that the structure of the last term implies that the tensor $\Pi_{\mu\nu}$ should be symmetric, whereas the first term implies that the tensor $\Phi_{\mu\nu\rho\sigma}$ should be constructed with the following symmetry, $\Phi_{\mu\nu\rho\sigma}=\Phi_{\rho\sigma\mu\nu}$. Next, varying (\ref{14}) with respect to $u^{\mu}$ and then demanding that $\delta\mathcal{S}=0$ for any variation $\delta u^{\mu}$, vanishing on the boundaries, yields
\begin{align}\label{15}
\left(-2\nabla^{\nu}\Phi_{\nu\mu\rho\sigma}+\Psi_{\mu\rho\sigma}-\Psi_{\sigma\rho\mu}\right)\nabla^{\rho}u^{\sigma}
-2\Phi_{\nu\mu\rho\sigma}\nabla^{\nu}\nabla^{\rho}u^{\sigma}+\left(2\Pi_{\mu\nu}-\nabla^{\rho}\Psi_{\nu\rho\mu}\right)u^{\nu}=0.
\end{align}
As with the case discussed in the previous section, it is sufficient for the the above equation to be satisfied, without imposing any constraint on the vector field $u^{\mu}$, to have the content of the first and the last parenthesis, as well as the tensor $\Phi_{\mu\nu\rho\sigma}$, vanish identically and separately. A weaker constraint, however, may be obtained when the tensor $\Phi_{\mu\nu\rho\sigma}$ possesses the following antisymmetry, $\Phi_{\mu\nu\rho\sigma}=-\Phi_{\rho\nu\mu\sigma}$. Indeed, in that case, the second derivative of the vector field $u^{\nu}$ in the above second-order differential equation transforms into a linear algebraic term in the vector field thanks to the identity $2\nabla_{[\mu}\nabla_{\nu]}u^{\rho}=R^{\rho}_{\:\:\sigma\mu\nu}u^{\sigma}$. The term then combines with the content of the last parenthesis and we are left only with the requirement that the content of the two parentheses independently and identically vanish. This translates thereby into the following two differential equations
\begin{align}\label{16}
2\nabla^{\nu}\Phi_{\nu\mu\rho\sigma}-\Psi_{\mu\rho\sigma}+\Psi_{\sigma\rho\mu}&=0,\nonumber
\\2\Pi_{\mu\nu}-\nabla^{\rho}\Psi_{\nu\rho\mu}-{R_{\nu}}^{\tau\sigma\rho}\Phi_{\rho\mu\sigma\tau}&=0.
\end{align}

Now that we have obtained the necessary and sufficient requirements to have an extremal entropy for every displacement vector field $u^{\mu}$, we shall build the tensors $\Phi_{\mu\nu\rho\sigma}$, $\Psi_{\mu\nu\rho}$, and $\Pi_{\mu\nu}$ whose general forms will be subject to the above symmetries and the constraints (\ref{16}). It is at this point that one should decide about the order of approximation in $G_{N}$ one wishes to include inside the entropy functional before building those tensors. However, including the Riemann tensor $R_{\mu\nu\rho\sigma}$ as well as the Ricci tensor $R_{\mu\nu}$, presupposes that the minimum approximation level one wishes to achieve inside the functional is already greater than or equals $G^{0}_{N}$. In this section, we shall restrict ourselves to this order. But the pattern one should follow to obtain higher-order approximations should be clear and will be outlined in Sec.~\ref{sec:5}.

At this level of approximation, one should not allow in the tensor $\Phi_{\mu\nu\rho\sigma}$ the appearance of terms containing more than one Riemann tensor $R_{\mu\nu\rho\sigma}$ or more than one Ricci tensor $R_{\mu\nu}$. Likewise, $\Psi_{\mu\nu\rho}$ should not contain more than three derivatives. Then, we may, at most, only include all the possible contractions of $R_{\mu\nu\rho\sigma}$ as well as $R_{\mu\nu}g_{\rho\sigma}$ with $\nabla^{\mu}u^{\nu}\nabla^{\rho}u^{\sigma}$. We should also construct from these tensors and their gradients, as well as gradients of other scalar functionals, all possible objects with three indices to be contracted with $u^{\mu}\nabla^{\nu}u^{\rho}$. However, as with the passage from integral (\ref{5}) to its generalized version (\ref{6}), one should allow, for completeness, each of the coupled terms to be weighted by a scalar field instead of putting a mere constant. Then, putting a hat above the scalars that have already been used in the construction (\ref{6}), the most general tensors we could build are of the following form
\begin{equation}\label{17}
\Pi_{\mu\nu}=\hat{\lambda} g_{\mu\nu}+T_{\mu\nu},
\end{equation}
\begin{align}\label{18}
\Phi_{\mu\nu\rho\sigma}=\;&\hat{A}\left(g_{\mu\sigma}g_{\nu\rho}-g_{\mu\nu}g_{\rho\sigma}\right)+A\left(R_{\mu\nu\rho\sigma}
-R_{\rho\nu\mu\sigma}\right)\nonumber
\\&+a\left(R_{\mu\nu}g_{\rho\sigma}+R_{\rho\sigma}g_{\mu\nu}-R_{\nu\rho}
g_{\mu\sigma}-R_{\mu\sigma}g_{\nu\rho}\right),
\end{align}
\begin{align}\label{19}
\Psi_{\mu\nu\rho}=\;&B_{,\sigma}{R^{\sigma}}_{\mu\nu\rho}+C_{,\sigma}{R^{\sigma}}_{\nu\mu\rho}
+\left(D_{,\sigma}{R^{\sigma}}_{\mu}+\hat{D}_{,\mu}\right)g_{\nu\rho}\nonumber
\\&+\left(E_{,\sigma}{R^{\sigma}}_{\nu}+\hat{E}_{,\nu}\right)g_{\mu\rho}+\left(F_{,\sigma}{R^{\sigma}}_{\rho}+\hat{F}_{,\rho}\right)g_{\mu\nu}\nonumber
\\&+I_{,\rho}R_{\mu\nu}+J_{,\nu}R_{\rho\mu}+K_{,\mu}R_{\nu\rho}+L{R^{\sigma}}_{\mu\nu\rho,\sigma}+M{R^{\sigma}}_{\nu\mu\rho,\sigma},
\end{align}
where $a$, $A$, ..., $M$ are new scalar functionals to be determined, while $T_{\mu\nu}$ is, as before, the energy-momentum tensor of matter. Note, first, that we put in front of $A$ a combination of two Riemann tensors instead of one to highlight its similarities with $\hat{A}$. Note, in addition, that all the previous terms from which emerged $\mathcal{F}(\phi,R)$-gravity are reintroduced here with a hat in order to keep track of the role of each additional term at each level of the construction. This will help us recognize the general pattern behind the whole approach as will be discussed in Sec.~\ref{sec:5}.

Also, one might actually still suspect at this order possible additional terms to be added to the tensor $\Psi_{\mu\nu\rho}$; terms such as  a third-order derivative of a scalar or the gradient of the Ricci tensor. Using the second contracted Bianchi identity, however, the last two terms inside $\Psi_{\mu\nu\rho}$ can in fact be brought to a combination of the derivatives $R_{\mu\nu,\rho}$. In addition, the first two terms, where the gradients of $B$ and $C$ are contracted with the Riemann tensor, may be written as third-order derivatives of these scalars. Similarly, adding any derivatives of scalars inside $\Pi_{\mu\nu}$ or inside $\Phi_{\mu\nu\rho\sigma}$ will be equivalent to just renaming one of the scalars or gradients already contained inside $\Psi_{\mu\nu\rho}$, as it follows, respectively, from the first and last parenthesis in (\ref{15}). The forms (\ref{17})-(\ref{19}) contain thereby the most general non-redundant terms one might include inside the entropy integral.

Now since the Riemann and the Ricci tensors already have the dimensions of $(\mathrm{lenght})^{-2}$, the functionals $A$ and $a$ must both be dimensionless, otherwise the approximation level inside the entropy functional would be greater than $G^{0}_{N}$. Likewise, since the tensor $\Psi_{\mu\nu\rho}$ is multiplied by one derivative of the field $u^{\mu}$, the functionals $B$, ..., $M$ may only depend on the scalar field $\phi$ but not on its derivatives or on the Ricci scalar $R$. In contrast, the hatted functionals $\hat{A}$, $\hat{D}$, $\hat{E}$, and $\hat{F}$ may contain not only the field $\phi$ but also its derivatives up to the second order, as well as the Ricci scalar $R$.

As we did in the previous section, we shall use the extremality constraints (\ref{16}) to find a relation between all these functionals and then deduce their general forms. Substituting the above expressions of $\Phi_{\mu\nu\rho\sigma}$ and $\Psi_{\mu\nu\rho}$ into the first constraint of (\ref{16}), we learn that our scalar functionals must satisfy the following differential equations
\begin{align}\label{20}
&R_{\mu\rho}\left(K-I-2a\right)_{,\nu}+{R^{\sigma}}_{\mu\rho\nu}\left(2A-B-2C\right)_{,\sigma}+R_{\rho\nu,\mu}\left(2A+2a-L-2M\right)\nonumber
\\&+\Big(\big[2\hat{A}+\hat{D}-\hat{F}+\Box(2a+F-D)\big]_{,\nu}-aR_{,\nu}-\Box\left[2a+F-D\right]_{,\nu}\Big)g_{\mu\rho}\nonumber
\\&-\{\mu\leftrightarrow\nu\}=0,
\end{align}
where $\{\mu\leftrightarrow\nu\}$ means the same sum as before but with the indicated indices everywhere interchanged. In deriving the above equation, we have used the first Bianchi identity ${R^{\mu}}_{[\nu\rho\sigma]}=0$, the contracted second Bianchi identity $\nabla^{\nu}R_{\nu\mu\rho\sigma}=2\nabla_{[\rho}R_{|\mu|\sigma]}$, as well as ${R^{\sigma}}_{\rho\mu\nu}\nabla^{\rho}f=2\nabla_{[\mu}\nabla_{\nu]}\nabla^{\sigma}f$, valid for any scalar $f$.

Before examining this equation, we turn to the second constraint in (\ref{16}). First, since the left-hand side contains both parities, we shall demand that the anti-symmetric and the symmetric parts vanish separately. Extracting first the anti-symmetric part, that reads $\nabla^{\rho}\Psi_{\nu\rho\mu}+{R_{\nu}}^{\tau\sigma\rho}\Phi_{\rho\mu\sigma\tau}-\{\mu\leftrightarrow\nu\}=0$, we find that
\begin{align}\label{21}
&\frac{1}{2}R_{,\nu}\left(I-K\right)_{,\mu}+{R^{\rho}}_{\nu}\left(D-F+I-K\right)_{,\mu\rho}+{R^{\rho}}_{\nu,\mu}\left(2C+B+D-F-L-2M\right)_{,\rho}
\nonumber\\&-\{\mu\leftrightarrow\nu\}=0.
\end{align}
In writing the above equation, we have used again the above Bianchi identities, the symmetry properties of the Riemann tensor, and the identity $\nabla^{\mu}R_{\mu\nu}=\frac{1}{2}\nabla_{\nu}R$.

Now, both differential equations (\ref{20}) and (\ref{21}) must be satisfied for any spacetime since the chosen form of the scalar functionals must be valid for any background geometry, regardless of its dynamics. Therefore, each factor in front of the different categories of tensors constituting the sums in these constraints must vanish identically and separately. Beginning with the identity (\ref{21}), we see that the vanishing of the factor multiplying $R_{,\nu}$ imposes to have $I-K=\mathrm{const}$. But, since both scalars appear differentiated once inside the tensor $\Psi_{\mu\nu\rho}$, this is equivalent to having $I=K$. With this, the vanishing of the factor in front of ${R^{\rho}}_{\nu}$ implies that $D-F=\mathrm{const}$ which, in turn, is equivalent to $D=F$ because again only gradients of these scalars are relevant inside the tensor $\Psi_{\mu\nu\rho}$. All that remains from (\ref{21}) then, is the following algebraic constraint
\begin{equation}\label{22}
2C+B-L-2M=\mathrm{const},
\end{equation}
where $\mathrm{const}$ is an arbitrary constant of integration that may be put to zero as we will see below. Going back now to (\ref{20}), we first see that the vanishing of the factor multiplying $R_{\mu\rho}$ implies, since we already have $K=I$, that $a=\mathrm{const}$. Furthermore, the vanishing of the factors in front of ${R^{\sigma}}_{\mu\rho\nu}$ and $R_{\rho\nu,\mu}$ yields, respectively,
\begin{align}\label{23}
2A-B-2C&=\mathrm{const},\nonumber
\\2A+2a-L-2M&=0,
\end{align}
where we have introduced another arbitrary integration constant that will be determined below. Finally, the remainder of (\ref{20}) gives the following first-order differential equation: $(2\hat{A}+\hat{D}-\hat{F})_{,\nu}=aR_{,\nu}$. Given that $a$ is found to be constant, this integrates to
\begin{equation}\label{24}
2\hat{A}(\phi,R)=aR-\left(\hat{D}-\hat{F}\right)(\phi,R),
\end{equation}
where we have indicated explicitly the eventual dependence of the hatted scalars upon the Ricci scalar $R$ and a scalar field $\phi$, and we have absorbed any integration constant inside $\hat{D}$ and $\hat{F}$ since both only appear differentiated inside the tensor $\Psi_{\mu\nu\rho}$.
With these preliminary results at hand, we may go back now to the second equation in (\ref{16}) and extract its symmetric part. The following field equations then result:
\begin{align}\label{25}
2T_{\mu\nu}=\;&R_{\mu\nu,\rho}(B+J+L)^{,\rho}+L\Box R_{\mu\nu}-L{R_{\nu\rho,\mu}}^{\rho}\nonumber
\\&+R_{\rho(\mu,\nu)}(2D-B-L)^{,\rho}+2{R^{\rho}}_{(\mu}(D+I)_{,\nu)\rho}\nonumber
\\&+g_{\mu\nu}(\Box\hat{E}+E_{,\rho\sigma}R^{\rho\sigma}+\frac{1}{2}E^{,\rho}R_{,\rho}-2\hat{\lambda})\nonumber
\\&+(\hat{D}+\hat{F})_{,\mu\nu}+I_{,(\mu}R_{,\nu)}+R_{\mu\nu}(2\hat{A}+\Box J)\nonumber
\\&-2aR_{\rho\mu}{R^{\rho}}_{\nu}+R_{\mu\rho\nu\sigma}(B^{,\rho\sigma}-2aR^{\rho\sigma})-AR_{\mu\rho\sigma\tau}{R_{\nu}}^{\rho\sigma\tau}.
\end{align}
Al that remains now, is to apply to this equality the conservation equation of the energy-momentum tensor of matter. After using the fact that $a$ is constant and that $\nabla^{\rho}\nabla_{\mu}R_{\nu\rho}=\nabla^{\rho}\nabla_{\nu}R_{\mu\rho}$, as well as the above mentioned Bianchi identities, we find,
\begin{align}\label{26}
&\frac{1}{2}{R_{,\mu}}^{\mu}I_{,\nu}-\frac{1}{2}{R_{,\nu}}^{\mu}\left(D+I\right)_{,\mu}+R^{\rho\mu}\Big(D+I\Big)_{,\nu\rho\mu}\nonumber
\\&+\frac{1}{2}R_{\rho\mu,\nu}\Big(2D+B-L\Big)^{,\mu\rho}+R_{\nu\sigma\mu\rho}\Big(2A+B\Big)^{,\rho\sigma\mu}\nonumber
\\&+\frac{1}{2}R_{\mu\nu,\rho}\Big(4D+2J+L-B+2I\Big)^{,\mu\rho}\nonumber
\\&+\frac{1}{2}{R_{\mu\nu,\rho}}^{\mu}\Big(2D+2J+B-L\Big)^{,\rho}\nonumber
\\&+\frac{1}{2}{R_{\rho\nu,\mu}}^{\mu}\Big(2D-B+L\Big)^{,\rho}\nonumber
\\&+{R_{\mu\nu,\rho}}^{\rho\mu}\Big(2A+2a+L\Big)-{R_{\rho\nu,\mu}}^{\rho\mu}\Big(2A+2a+L\Big)\nonumber
\\&+R_{\rho\nu}\Big(\big[2\hat{A}+\hat{D}+\hat{F}+\Box J\big]^{,\rho}-aR^{,\rho}+\Box\big[D+I\big]^{,\sigma}\Big)\nonumber
\\&+\Big(\Box[\hat{D}+\hat{E}+\hat{F}]+E_{,\mu\rho}R^{\mu\rho}+\frac{1}{2}R_{,\mu}E^{,\mu}-2\hat{\lambda}\Big)_{,\nu}\nonumber
\\&+\frac{1}{2}\left(R_{,\mu}[D+2I]^{,\mu}\right)_{,\nu}+\Big(\frac{1}{2}\Box[J+I]+\hat{A}\Big)R_{,\nu}\nonumber
\\&-a\left(R_{\rho\mu}R^{\rho\mu}\right)_{,\nu}
-\frac{1}{4}A\left(R_{\mu\rho\sigma\tau}R^{\mu\rho\sigma\tau}\right)_{,\nu}=0.
\end{align}
In rearranging the terms to achieve the above expression, we have also used the identity ${R^{\mu}}_{\rho}\nabla_{\mu}{R^{\rho}}_{\nu}=\frac{1}{2}\nabla_{\nu}\left(R_{\rho\mu}R^{\rho\mu}\right)+R^{\rho\sigma}\nabla^{\mu}R_{\mu\rho\sigma\nu}$.

A reasoning similar to that conducted in the paragraph below equation (\ref{21}) implies that, except for the last three rows, each factor contracting each category of curvature tensors in each row should vanish independently in (\ref{26}). Beginning with the first term in the first row, we learn that $I=\mathrm{const}$, but since this scalar appears inside the entropy integral differentiated once, the constant is meaningless and the result is equivalent to $I=0$. The vanishing of the factor contracting ${R_{,\nu}}^{\mu}$ in the same row also implies that $D+I=\mathrm{const}$, but since only the gradient of $D$ is relevant, this is equivalent to having $D=0$. This, in turn, makes the remaining terms of that row vanish identically.

Next, the independently vanishing factors contracting the tensors $R_{\rho\mu,\nu}$ and $R_{\mu\nu,\rho}$ in the second and the third row imply that $B-L=\mathrm{const}$ and $2J+L-B=\mathrm{const}$, respectively, imposing thereby to have $J=\mathrm{const}$. Again, since $J$ is differentiated inside the entropy integral, this is equivalent to having $J=0$. Hence, $B=L+\mathrm{const}$. Given that only the gradient of $B$ matters, this is equivalent to $B=L$. Consequently, both contracting factors in front of ${R_{\mu\nu,\rho}}^{\mu}$ and ${R_{\rho\nu,\mu}}^{\mu}$ in the fourth and the fifth row, respectively, automatically vanish. Furthermore, with $B=L$ the constraint (\ref{22}) gives $C=M$, where the constant of integration obtained there is absorbed inside the scalar $C$ since only the gradient of the latter is relevant.

With these results, the vanishing of the sixth row in (\ref{26}) yields a single constraint, namely $A=-a-L/2$. This, together with the second constraint in (\ref{23}), imply that $L=-M$. Altogether then, we have $B=-C=-M=L$ and $A=-a+M/2$. Referring to the first constraint in (\ref{23}), we learn that the constant of integration there is nothing else than $-2a$, which is thus the only arbitrary constant that arises from the whole formalism.

All that remains then in (\ref{26}) is the seventh row, as well as the whole content of the last three rows. The vanishing of the former gives, after setting $D=I=J=0$, the following differential equation, $(2\hat{A}+\hat{D}+\hat{F})_{,\nu}=aR_{,\nu}$, which integrates to
\begin{equation}\label{27}
2\hat{A}(\phi,R)=aR-\left(\hat{D}+\hat{F}\right)(\phi,R).
\end{equation}
Comparison with (\ref{24}), reveals that $\hat{F}=\mathrm{const}$, which is equivalent to $\hat{F}=0$ since once again only the gradient of this scalar is relevant inside the entropy integral. Finally, the vanishing of the content of the last three rows of (\ref{26}) gives
\begin{align}\label{28}
\Big(2\hat{\lambda}-\Box[\hat{D}+\hat{E}]-R_{\rho\mu}E^{,\rho\mu}-\frac{1}{2}R_{,\mu}E^{,\mu}\Big)_{,\nu}=
\hat{A}R_{,\nu}-a\left(R_{\rho\mu}R^{\rho\mu}\right)_{,\nu}-\frac{A}{4}\left(R_{\mu\rho\sigma\tau}R^{\mu\rho\sigma\tau}\right)_{,\nu}.
\end{align}
Setting the content of the parenthesis in the left-hand side of this equation equal to $\mathcal{F}$, where $\mathcal{F}$ is a scalar functional of the scalars $\phi$, $R$, $P$, and $Q$, we learn that
\begin{equation}\label{29}
\frac{\delta\mathcal{F}}{\delta\phi}=0,\qquad\frac{\partial\mathcal{F}}{\partial R}=\hat{A},\qquad\frac{\partial\mathcal{F}}{\partial P}=-a,\qquad\frac{\partial\mathcal{F}}{\partial Q}=-\frac{A}{4}.
\end{equation}
From the argument made in the second paragraph below (\ref{19}) we know that the scalar $\hat{D}$ may contain, at most, one positive power of $R$. Therefore, writing $\hat{A}=aR/2-\hat{D}/2$ as it follows from (\ref{27}) and the fact that $A=-a+M/2$, the last three partial differential equations in (\ref{29}) may be integrated to give
\begin{equation}\label{30}
\mathcal{F}=\frac{a}{4}\mathcal{R}^{2}_{GB}+\mathcal{M}(\phi)R_{\mu\nu\rho\sigma}R^{\mu\nu\rho\sigma}+\mathcal{D}(\phi,R)+U(\phi).
\end{equation}
Here, $\mathcal{R}^{2}_{GB}$ is the Gauss-Bonnet invariant, $\mathcal{M}(\phi)=-\frac{1}{8}M(\phi)$, the functional $\mathcal{D}(\phi,R)$ is at most quadratic in $R$ and given by $-\frac{1}{2}\int\hat{D}(\phi,R)\mathrm{d}R$, whereas $U(\phi)$ is some functional only of $\phi$ and its derivatives up to the second order.

Now from the last three identities in (\ref{29}) it follows that the relations $A=-a+M/2$ and $\hat{A}=aR/2-\hat{D}/2$ also read $M=-8\mathcal{F}_{Q}-2\mathcal{F}_{P}$ and $\hat{D}=-R\mathcal{F}_{P}-2\mathcal{F}_{R}$, respectively. Therefore, substituting these into the field equations (\ref{25}), and using the fact that $D=I=J=\hat{F}=0$ and $B=-C=-M=L$, we obtain
\begin{align}\label{31}
T_{\mu\nu}=\;&g_{\mu\nu}[-\frac{1}{2}\mathcal{F}+\Box\mathcal{F}_{R}+\frac{1}{2}\mathcal{F}_{P}\Box R]+R_{\mu\nu}\mathcal{F}_{R}\nonumber
\\&+2R_{\rho\mu}{R^{\rho}}_{\nu}\mathcal{F}_{P}+2R_{\mu\rho\sigma\tau}{R_{\nu}}^{\rho\sigma\tau}\mathcal{F}_{Q}
-{\mathcal{F}_{R}}_{,\mu\nu}\nonumber
\\&+\mathcal{F}_{P}\Box R_{\mu\nu}-4R_{\mu\rho\sigma\nu}{\mathcal{F}_{Q}}^{,\rho\sigma}+8R_{\mu\nu,\rho}{\mathcal{F}_{Q}}^{,\rho}\nonumber
\\&+4\mathcal{F}_{Q}\Box R_{\mu\nu}-4{R_{\mu\rho,\nu}}^{\rho}\mathcal{F}_{P}
-8R_{\rho(\mu,\nu)}{\mathcal{F}_{Q}}^{,\rho}\nonumber
\\&-2{R_{\mu\rho,\nu}}^{\rho}\mathcal{F}_{P}.
\end{align}
In writing the above equation we have used the identity $R^{\rho\sigma}R_{\mu\rho\sigma\nu}=\nabla^{\rho}\nabla_{\nu}R_{\mu\rho}-\frac{1}{2}\nabla_{\mu}\nabla_{\nu}R-R_{\mu\rho}{R^{\rho}}_{\nu}$, thanks to which certain terms combine among themselves while others cancel each other. Now, referring to the field equations (\ref{4}) it becomes evident, after recalling that $\mathcal{F}_{P}$ is constant in our case and using also the identity $\nabla^{\rho}\nabla^{\sigma}R_{\mu\rho\sigma\nu}=\nabla^{\rho}\nabla_{\mu}R_{\nu\rho}-\Box R_{\mu\nu}$, that (\ref{31}), together with the result $\delta\mathcal{F}/\delta\phi=0$, are nothing but the field equations of an $\mathcal{F}(\phi,R,P,Q)$-modified gravity, where the functional $\mathcal{F}$ is given by (\ref{30}).

From this result, it follows that the extended entropy functional formalism developed here implies that, contrary to what one might expect, at the second order in the curvature the terms $\Box R$ and $\partial^{\mu}\phi\partial_{\mu}R$ do not arise inside the gravitational action. Furthermore, one does not even obtain a scalar-Gauss-Bonnet gravity, but the pure GB term plus a scalar field coupled to the Ricci scalar $R$ and its square $R^{2}$, as well as a coupling between the scalar field and the Kretschmann scalar $R_{\mu\nu\rho\sigma}R^{\mu\nu\rho\sigma}$. The latter appears specifically in the effective action of the heterotic-type I strings (see e.g. \cite{Heterotic} for a discussion of the relevance of the latter coupling in the AdS$_{5}$ supergravity). The functional $U(\phi)$ in (\ref{30}) would then simply contain the scalar field's potential as well as its kinetic energy terms.

\section{Towards higher-order corrections}\label{sec:5}
As we saw in Sec.~(\ref{sec:3}), $\mathcal{F}(\phi,R)$-modified gravity wouldn't have been possible to obtain from the entropy functional had we restrained the latter to contain couplings between the metric tensor $g_{\mu\nu}$ and the quadratic terms $(\nabla u)^{2}$ weighted only by constant multiplicative factors. Indeed, as we discussed in the last paragraph of Sec.~\ref{sec:3}, the scalar field gradients wouldn't also appear and no coupling with the hybrid terms $u\nabla u$ would have arisen. It is because of these latter terms that one can go from General Relativity to an $\mathcal{F}(\phi,R)$-modified gravity, and from the Gauss-Bonnet gravity to an $\mathcal{F}(\phi,R,Q)$-Gauss-Bonnet gravity as we saw in the preceding section. The same pattern actually repeats itself each time one attempts to achieve higher-order approximations inside the entropy integral (\ref{14}).

One begins by introducing, in accordance with the desired order of approximation, the right number of contracted Riemann and Ricci tensors to couple with the quadratic terms $(\nabla u)^{2}$. One then weights each term with a spacetime-dependent factor, that is, a scalar functional of the curvature tensors, the scalars $R$, $P$, $Q$,..., and/or an independent scalar field and their derivatives. At each new order, thanks to the symmetries of the tensor $\Phi_{\mu\nu\rho\sigma}$, the latter keeps growing in the number of the Riemann and the Ricci tensors it contains, giving rise to the corresponding Lanczos-Lovelock Lagrangians as already found and exposed in detail in \cite{Pad+Par}.

However, because of the hybrid terms $u\nabla u$, the scalar functionals weighting each term may not all be constants. Therefore, when going to higher-order approximations inside the integral (\ref{14}), one will recover the entire series of Lanczos-Lovelock Lagrangians but accompanied at each order in the curvature with specific additional terms featuring non-minimal couplings of the scalar field with the background's geometric tensors. The precise structure of each term would be dictated by the constraints (\ref{16}). In fact, at each level, one just recovers the structure that appeared in the previous level (the terms denoted with a hat, as we saw) plus higher products of the Riemann and Ricci tensors and their derivatives.

While we have not yet elaborated a general procedure for a systematic investigation of the higher orders in the curvature (which will be attempted in a separate paper), one might already expect to recover each order of the Lanczos-Lovelock Lagrangian augmented with non-trivial couplings between specific curvature tensors and the field $\phi$.

From this discussion, it appears then that this approach becomes limitless and helps obtain the precise modifications necessary to be brought to the Einstein-Hilbert action at a given order in the curvature, just by including the relevant orders in $\mathrm{G}$ inside the entropy functional.

\section{Block hole entropy in the extended formalism}\label{sec:6}
Now that we found what structure the quadratic curvature corrections must possess within the framework of this extended entropy functional formalism, the next natural task is to apply the resulting extremal functional (called the 'on-shell' entropy \cite{Pad,Pad+Par}) to analyze the entropy of black holes in the induced modified gravity theories. So let us first go back to integral (\ref{14}) and perform an integration by parts, taking into account the constraints (\ref{16}), that is, the extremalization conditions. We find the following general 'on-shell' functional
\begin{equation}\label{32}
{\cal S}=\int_{\partial \mathcal{V}}\mathrm{d}^{3}x\sqrt{|h|}n^{\mu}\left(\Phi_{\mu\nu\rho\sigma}u^{\nu}\nabla^{\rho}u^{\sigma}
+\frac{1}{2}\Psi_{\nu\mu\rho}u^{\nu}u^{\rho}\right),
\end{equation}
where $h$ is the determinant of the induced three-metric on the boundary $\partial\mathcal{V}$. This result will be valid at any order in the curvature achieved inside the tensors $\Phi_{\mu\nu\rho\sigma}$ and $\Psi_{\mu\nu\rho}$.

The general procedure for dealing with Killing horizons using the entropy functional formalism is explained in \cite{Pad,Pad+Par,Pad2}, and a detailed calculation using the local Rindler frame can be found there. So, here we shall just briefly give an outline of the approach and then apply it to integral (\ref{32}).

When using spherical coordinates, the procedure consists in choosing first a static spherical symmetric metric $\mathrm{d}s^{2}=-f(r)\mathrm{d}t^{2}+\mathrm{d}r^{2}/f(r)+r^{2}\mathrm{d}\Omega^{2}$, where $\mathrm{d}\Omega^{2}=\mathrm{d}\theta^{2}+\sin^{2}\theta\mathrm{d}\phi^{2}$, and $f(r)$ is a smooth function such that $f(r_{\mathcal{H}})=0$; making the surface $r=r_{\mathcal{H}}$ a Killing horizon $\mathcal{H}$. Next, one identifies the displacement vector field $u^{\mu}$ with the unit space-like normal to the surfaces $r=\mathrm{const}$. Hence, one chooses the vector $u^{\mu}$ to be $u^{\mu}=(0,\sqrt{f(r)},0,0)=n^{\mu}$. Further, when performing the time integral one restricts the time variable to the range $[0,2\pi/\kappa]$, i.e., one integrates over a periodic time, where $\kappa=\frac{1}{2}\partial_{r}f|_{\mathcal{H}}$ is the surface gravity at the Killing horizon $r=r_{\mathcal{H}}$. Finally, one must insert these ingredients inside the entropy integral and take the limit $r\rightarrow r_{\mathcal{H}}$ at the very end of the calculation.

Substituting the above chosen expression for $u^{\mu}$ and computing the covariant derivative $\nabla^{\rho}u^{\sigma}$ using the above spherical metric, we get, after making use of the antisymmetry $\Phi_{\mu\nu\rho\sigma}=-\Phi_{\rho\nu\mu\sigma}$,
\begin{align}\label{33}
{\cal S}=\lim_{r\rightarrow r_{\mathcal{H}}}\int^{2\pi/\kappa}_{0}\mathrm{d}t\int\mathrm{d}\theta\mathrm{d}\phi
\Big(\frac{\partial_{r}f}{2}{\Phi^{0}}_{110}+\frac{f}{r}{\Phi^{2}}_{112}+\frac{f}{r}{\Phi^{3}}_{113}+\frac{f}{2}\Psi_{111}\Big)r^{2}f\sin\theta.
\end{align}
From (\ref{18}) and (\ref{19}), we find for $i=0,2,3$, respectively,
\begin{align}\label{34}
{\Phi^{i}}_{11i}&=\frac{\hat{A}}{f}+A{R^{i}}_{11i}-\frac{a}{f}{R^{1}}_{1}-\frac{a}{f}{R^{i}}_{i},\nonumber
\\\Psi_{111}&=\frac{1}{f}\left(\partial_{r}\hat{D}+\partial_{r}\hat{E}+{R^{1}}_{1}\partial_{r}E\right).
\end{align}
Computation of the components ${R^{i}}_{11i}$ of the Riemann tensor and ${R^{i}}_{i}$ of the Ricci tensor corresponding to our spherical symmetric metric gives the following values
\begin{align}
&{R^{0}}_{110}=\frac{\partial^{2}_{r}f}{2f},\quad{R^{2}}_{112}={R^{3}}_{113}=\frac{\partial_{r}f}{2rf},\nonumber
\\&{R^{0}}_{0}={R^{1}}_{1}=-\frac{\partial^{2}_{r}f}{2}-\frac{\partial_{r}f}{r},\quad
{R^{2}}_{2}={R^{3}}_{3}=\frac{1}{r^{2}}-\frac{f}{r^{2}}-\frac{\partial_{r}f}{r}.\nonumber
\end{align}
Inserting these values in (\ref{34}) reveals that all the terms inside the parenthesis in integral (\ref{33}) have a finite limit at $r\rightarrow r_{\mathcal{H}}$ except for the term ${\Phi^{0}}_{110}$, in which $f$ appears in some of its denominators. Indeed, writing $\hat{A}$ and $A$ in terms of the functionals $\mathcal{D}$ and $\mathcal{M}$ appearing inside the expression of $\mathcal{F}$ in (\ref{30}) as
\begin{align}\label{35}
\hat{A}=\frac{1}{2}aR+\frac{\partial\mathcal{D}}{\partial R},\quad A=-a-4\mathcal{M},
\end{align}
we find from the first identity in (\ref{34}) that
\begin{equation}\label{36}
{\Phi^{0}}_{110}=-\frac{a}{r^{2}}+\frac{a}{r^{2}f}+\frac{1}{f}\frac{\partial\mathcal{D}}{\partial R}-2\mathcal{M}\frac{\partial^{2}_{r}f}{f}.
\end{equation}
Inserting this into (\ref{33}) and taking the limit $r\rightarrow r_{\mathcal{H}}$, gives
\begin{equation}\label{37}
{\cal S}=2\pi\mathcal{A}\left(\frac{a}{r^{2}_{\mathcal{H}}}+\frac{\partial\mathcal{D}}{\partial R}\Big|_{\mathcal{H}}-2\mathcal{M}\partial^{2}_{r}f\Big|_{\mathcal{H}}\right),
\end{equation}
where $\mathcal{A}$ is the surface area of the horizon and the subscript $\mathcal{H}$ means that the corresponding quantities are evaluated on the horizon.

First of all, we note that when setting $a=\mathcal{M}=0$ and $\mathcal{D}=R/8\pi G_{N}$, we satisfactorily recover exactly the black hole entropy $\mathcal{S}=\mathcal{A}/4G_{N}$ of General Relativity. Actually, in this special case the scalar $\hat{A}$ becomes the constant $1/8\pi G_{N}$, as it follows from (\ref{35}), whereas all the other terms in the expressions (\ref{18}) and (\ref{19}) vanish. The entropy functional (\ref{14}) reduces thereby to the form (\ref{5}), with $B=0$ and $A=-C=\mathrm{const}=1/8\pi G_{N}$ there, which, as we saw, yields Einstein's field equations.

On the other hand, the first term of the more general case (\ref{37}) is what one obtains for the black hole entropy in a Gauss-Bonnet gravity. See e.g., \cite{f(GB)+Noether} where the result was obtained using the Noether charge method \cite{Noether}, and \cite{GB+Carlip} where the Carlip method \cite{Carlip} was used. The second term in (\ref{37}) is just the contribution to the black hole entropy in a $\mathcal{D}(\phi,R)$-modified gravity; see e.g., \cite{f(R)+Noether} where the result was obtained using the Noether charge method. The third term represents then the separate contribution to the black hole's entropy of the Kretschmann scalar. In the next section we will discuss in more detail this coincidence in the results obtained from our approach with those obtained in the literature from the Noether charge method using Wald's entropy.

\section{The new features in the extended formalism}\label{sec:7}
The original entropy functional formalism introduced by Padmanabhan et al. was based on the crucial insight \cite{Pad2,Pad3} that the field equations $2\mathcal{G}_{\mu\nu}=T_{\mu\nu}$ of any invariant theory $\mathcal{L}$ under the diffeomorphism $x^{\mu}\rightarrow x^{\mu}+u^{\mu}$, where
\begin{equation}\label{38}
\mathcal{G}^{\mu\nu}=\frac{\partial\mathcal{L}}{\partial R_{\mu\rho\sigma\delta}}R^{\nu}_{\rho\sigma\delta}-\frac{1}{2}\mathcal{L}g^{\mu\nu}-2\nabla_{\rho}\nabla_{\sigma}\frac{\partial\mathcal{L}}{\partial R_{\mu\rho\sigma\nu}},
\end{equation}
can be obtained by demanding that the integral $-\int\mathrm{d}^{4}x\sqrt{-g}(2\mathcal{G}_{\mu\nu}-T_{\mu\nu})n^{\mu}n^{\nu}$ be extremal for all null vector fields $n^{\mu}$. Thus, by performing an integration by parts on the term $\mathcal{G}_{\mu\nu}$ after using (\ref{38}), the approach amounts to imposing an extremality condition on the following integral \cite{Pad+Par,Pad2}
\begin{equation}\label{39}
-4\int\mathrm{d}^{4}x\sqrt{-g}\big[\mathcal{P}^{\mu\nu\rho\sigma}\nabla_{\rho}n_{\mu}\nabla_{\sigma}n_{\nu}
+\left(\nabla_{\sigma}\mathcal{P}^{\mu\nu\rho\sigma}\right)n_{\nu}\nabla_{\rho}n_{\mu}
+(\nabla_{\rho}\nabla_{\sigma}\mathcal{P}^{\mu\rho\sigma\nu}-\frac{T_{\mu\nu}}{4})n_{\mu}n_{\nu}\big],
\end{equation}
where $\mathcal{P}^{\mu\nu\rho\sigma}=\partial\mathcal{L}/\partial R_{\mu\nu\rho\sigma}$. On the other hand, it was shown in Refs.~\cite{Pad+Par,Pad2} that the latter integral provides an elegant physical interpretation when spacetime is viewed as a continuous elastic medium subject to the deformation $u^{\mu}$; it would simply represent the entropy of null surfaces expressed in terms of the coarse grained degrees of freedom of spacetime that underlies the microscopic degrees of freedom of the latter. Therefore, since Wald's horizon entropy for any diffeomorphism invariant theory is $2\pi/\kappa$ times the Noether charge associated with diffeomorphism invariance and produced by the Noether current flux density $n_{\mu}\mathcal{J}^{\mu}=2n_{\mu}n_{\nu}\mathcal{G}^{\mu\nu}$\cite{Noether}, this approach provides a novel variational principle for gravity deeply rooted in thermodynamics. Indeed, the above integral measures simply the balance between the gravitational entropy current density $2\pi\mathcal{J}^{\mu}/\kappa$ and the matter entropy current density $2\pi T^{\mu\nu}n_{\nu}/\kappa$ across a horizon whose null normal is $n^{\mu}$ \cite{Pad+Par,Pad2}. It is this relation to Wald's entropy that makes it possible to recover within Padmanabhan's entropy functional formalism the same results one obtains when using the Noether charge method.

In our present extended formalism, however, we have generalized the approach further by keeping only the basic idea that consists in extremizing an entropy functional in accordance with the second principle of thermodynamics, as well as the interpretation of the field $u^{\mu}$ inside the functional as a displacement vector field. Indeed, in contrast to (\ref{39}), we did not choose or rely on any a priori structure of the two functionals $\Phi_{\mu\nu\rho\sigma}$ and $\Psi_{\mu\nu\rho}$ appearing inside our entropy integral (\ref{14}). Instead, we found the structures of these functionals from the variational principle itself by demanding that the latter be satisfied for \textit{any} displacement vector field $u^{\mu}$. It is precisely this last point that makes our approach capable -- thanks to the system (\ref{16}) -- of imposing nontrivial constraints on the structure of the higher-order modified gravity theories one might build from the formalism. In fact, as it was also shown in Refs.~\cite{WWGY,Pad3}, when one generalizes the approach by choosing right from the outset the structure (\ref{39}) for the entropy functional \footnote{This choice was justified in Ref.\cite{WWGY} by invoking an interesting analogy with the free energy of continuous media in which one encounters varying elastic and piezoelectric 'constants', and electric permittivity.}, one merely recovers the general form of the equations of motion of a diffeomorphism invariant theory. Indeed, when performing an integration by parts inside integral (\ref{39}) with a null vector field $n^{\mu}$, the integral becomes (up to a term proportional to $g_{\mu\nu}n^{\mu}n^{\nu}$ inside the first parentheses \cite{WWGY,Pad3})
\begin{align}\label{40}
-\int_{\mathcal{V}}\mathrm{d}^{4}V\left(2\mathcal{P}^{\mu\rho\sigma\delta}R^{\nu}_{\rho\sigma\delta}
-4\nabla_{\rho}\nabla_{\sigma}\mathcal{P}^{\mu\rho\sigma\nu}-T^{\mu\nu}\right)n_{\mu}n_{\nu}
-4\int_{\partial\mathcal{V}}\mathrm{d}^{3}\Sigma_{\sigma}\mathcal{P}^{\mu\nu\rho\sigma}n_{\nu}\nabla_{\rho}n_{\mu},
\end{align}
where $\mathrm{d}^{4}V$ is the invariant four-volume element, whereas $\mathrm{d}^{3}\Sigma_{\sigma}$ is the three-surface element whose normal is in the direction $n_{\sigma}$. This latter expression clearly displays the general structure of the field equations of diffeomorphism invariant theories, that would emerge from the bulk whenever (\ref{40}) is varied with respect to $n^{\mu}$, plus a boundary term which provides the same entropy of horizons one finds when using Wald's entropy (see, however, Ref.~\cite{WWGY} for an elaborate analysis on this last point).

In the light of the above discussion, the fact that the black hole entropy we found using formula (\ref{32}), deduced from our extended formalism (\ref{14}), coincides with the result one obtains using Wald's entropy might thereby appear as a mere coincidence. In what follows, however, we will see that this is far from being a coincidence because it actually has a deeper origin.

In order to expose more clearly the relation our approach bears with Wald's entropy, we will use our previous results for $\Phi_{\mu\nu\rho\sigma}$ and $\Psi_{\mu\nu\rho}$ at the second-order approximation in the curvature to rewrite our 'on-shell' entropy formula (\ref{32}) in terms of the tensor $\mathcal{P}^{\mu\nu\rho\sigma}$; the latter being given within the conventions used in Sec.~\ref{sec:2} by $\partial\mathcal{F}/2\partial R_{\mu\nu\rho\sigma}$. First, substituting identities (\ref{29}) together with $M=-8\mathcal{F}_{Q}-2\mathcal{F}_{P}$, $\hat{D}=-R\mathcal{F}_{P}-2\mathcal{F}_{R}$, $D=F=I=J=K=\hat{F}=0$ and $B=-C=-M=L$ into (\ref{18}) and (\ref{19}), we find, respectively,
\begin{equation}\label{41}
\Phi_{\mu\nu\rho\sigma}=\mathcal{F}_{R}\left(g_{\mu\sigma}g_{\nu\rho}-g_{\mu\nu}g_{\rho\sigma}\right)+4\mathcal{F}_{Q}R_{\rho\mu\nu\sigma}
-\mathcal{F}_{P}\left(R_{\mu\nu}g_{\rho\sigma}-R_{\nu\rho}g_{\mu\sigma}+R_{\rho\sigma}g_{\mu\nu}-R_{\mu\sigma}g_{\nu\rho}\right),
\end{equation}
\vspace{-0.5cm}
\begin{equation}\label{42}
\Psi_{\mu\nu\rho}=\nabla_{\sigma}\left[\left(8\mathcal{F}_{Q}+2F_{P}\right)R^{\sigma}_{\:\:\rho\nu\mu}\right]
-g_{\nu\rho}\nabla_{\mu}\left(R\mathcal{F}_{P}+2\mathcal{F}_{R}\right)+g_{\mu\rho}\left(R^{\sigma}_{\:\:\nu}\nabla_{\sigma}E+\nabla_{\nu}\hat{E}\right).
\end{equation}
On the other hand, from expression (\ref{30}) of the Lagrangian functional $\mathcal{F}(\phi,R,P,Q)$ we obtained in Sec.~\ref{sec:4}, we also easily deduce the following expression for the tensor $\mathcal{P}^{\mu\nu\rho\sigma}$:
\begin{align}\label{43}
\mathcal{P}^{\mu\nu\rho\sigma}=\frac{\partial\mathcal{F}}{2\partial R_{\mu\nu\rho\sigma}}=&\;\frac{1}{4}\mathcal{F}_{R}\left(g^{\mu\rho}g^{\nu\sigma}-g^{\mu\sigma}g^{\nu\rho}\right)
+\mathcal{F}_{Q}R^{\mu\nu\rho\sigma}\nonumber
\\&+\frac{1}{4}\mathcal{F}_{P}\left(R^{\mu\rho}g^{\nu\sigma}-R^{\nu\rho}g^{\mu\sigma}+R^{\nu\sigma}g^{\mu\rho}-R^{\mu\sigma}g^{\nu\rho}\right).
\end{align}
Hence, comparing (\ref{41}) and (\ref{43}) we learn that $\Phi_{\mu\nu\rho\sigma}=-4\mathcal{P}_{\mu\rho\nu\sigma}=-4\mathcal{P}_{\sigma\nu\rho\mu}$, where the last equality comes from the symmetries of the Riemann tensor. As for the tensor $\Psi_{\mu\nu\rho}$, we see by comparing (\ref{42}) and (\ref{43}) that it has no such simple expression in terms of the tensor $\mathcal{P}^{\mu\nu\rho\sigma}$.
Therefore, our 'on-shell' entropy integral (\ref{32}) may, sufficiently for our purposes, be written in terms of the tensor $\mathcal{P}^{\mu\nu\rho\sigma}$ as follows
\begin{align}\label{44}
{\cal S}=\int_{\partial \mathcal{V}}\mathrm{d}^{3}x\sqrt{|h|}n^{\mu}\left(-4\mathcal{P}_{\sigma\nu\rho\mu}u^{\nu}\nabla^{\rho}u^{\sigma}
+\frac{1}{2}\Psi_{\nu\mu\rho}u^{\nu}u^{\rho}\right).
\end{align}

The first term inside the parentheses in this integral has the same structure as the integrand of the surface contribution in (\ref{40}) found using the formalism (\ref{39}), whereas the second term inside the parentheses did not appear in the latter formalism. As we discussed it in Sec.~\ref{5} and at the end of Sec.~\ref{3}, though, it is thanks to this additional term that one might recover the more general modified gravity theories. Furthermore, as we saw in Sec.~\ref{6}, when one evaluates (\ref{44}) on the horizon, in which case the space-like displacement vector field normal to the stretched horizon \cite{Pad+Par,Pad2,WWGY} becomes the null normal to the true horizon, one is indeed left only with the first term which then coincides perfectly with the boundary term of (\ref{40}).

Hence, we now see exactly why Wald's entropy is recovered from our extended formalism. Wald's approach is recovered as a necessary built-in package that accompanies the equations of motion of every diffeomorphism invariant theory that emerges from the new variational principle. In other words, instead of starting from the celebrated relation between thermodynamics of horizons and the Noether charge of diffeomorphism invariant theories, our formalism gives naturally rise to diffeomorphism invariant theories for the spacetime background and simultaneously supplies us with the right horizon thermodynamics these theories would provide through the Noether charge associated with their diffeomorphism invariance. All this comes out from the single requirement that the functional be extremal for \textit{every} configuration of the displacement vector field $u^{\mu}$.

Finally, another peculiar feature worth discussing here is the following. It is well-known in classical mechanics that, whenever friction is present in a system, the equations of motion of the latter are not invariant under time-reversal. Therefore, having 'friction' terms inside the entropy functional would suggest at first sight that the equations of motion one would obtain for the emerging gravity would also not be invariant under time-reversal. However, as we saw in Sec.~\ref{sec:4}, the field equations obtained are invariant under time-reversal as are those of any diffeomorphism invariant theory of gravity. The reason for this goes back to the fact, already stressed in \cite{Pad,Pad+Par,Pad2}, that the variational principle used here is not intended to find the equations of motion of the displacement vector field $u^{\mu}$, as it is done in standard continuum mechanics, but rather to impose constraints on the background geometry of the medium for arbitrary configurations of the displacement vector field $u^{\mu}$. What one finds then are equations of motion for the background metric of spacetime that are indeed invariant under time-reversal. A good analogy here would be to invoke the fact that although the equations governing the individual atoms and molecules inside a fluid are invariant under time-reversal, their collective behavior translates into the macroscopic phenomenon of friction which exhibits non-invariance under time-reversal. Thus, although the extended entropy formalism exhibits irreversibility, when used to find out the dynamics of spacetime, it yields equations that are invariant under time-reversal.

\section{Summary and discussion}\label{sec:8}
The aim of this work was to examine the possibility of extending the entropy functional formalism to obtain, besides the Lanczos-Lovelocke theories, other known extensions of General Relativity. In so doing, we saw that it is possible to recover every $\mathcal{F}(\phi,R)$-modified gravity theory but only specific second-order corrections in the curvature. The gravitational Lagrangian at this order is given by the functional (\ref{30}). As we saw, however, the construction of the entropy integral implies that adopting an $\mathcal{F}(\phi,R)$-gravity without including the Riemann and the Ricci tensors is unnatural within this approach.

This extended entropy functional formalism shows that the whole approach may in fact be used as a tool to explore the higher-curvature modifications to General Relativity. The whole approach is simply based on general covariance, the conservation of the energy-momentum tensor of matter, and some physical concepts borrowed from the physics of continuous media. The calculations are straightforward and may thereby easily be extend further to include spacetimes with torsion and recover the so-called $\mathcal{F}(T)$-modified gravity (see e.g. \cite{f(T)} for the relevance of the latter to cosmology). This would generalize the construction made in \cite{Hammad+Erratum} to yield specific modifications to Einstein-Cartan gravity.

As mentioned in Sec.~\ref{sec:3}, we would like to note again that the power counting used in the approach does not exclude having at each order of approximation terms of the form $R\Box^{-k}R$. These terms would give rise to what is known as nonlocal gravity \cite{Reviews}. Doing so, however, would also bring nonlocal terms inside the entropy integral. Hence, in order to achieve nonlocal gravity from our approach, one must first provide a justification for including nonlocal couplings with the displacement field $u^{\mu}$ inside its corresponding entropy.

As it already emerged from the works of Padmanabhan et al., the entropy functional approach has also the merit of shedding more light on the holographic nature of gravity and the thermodynamics of null surfaces. In addition, as we saw, even black hole entropy in modified gravity came out right in the extended formalism. Furthermore, although in the latter, one does not impose any a priori constraint on the different parts constituting the entropy functional, one is remarkably led, thanks to the variational principle alone, to the same structure one obtains using Wald's approach based on the Noether charge of diffeomorphism invariant theories. Therefore, being of a thermodynamical origin, this formalism may be taken as a fundamental paradigm when it comes to searching for higher-energy corrections to General Relativity. In this sense, the power-counting in Newton's constant $G_{N}$ used here inside the entropy functional would be the analogue of an $\alpha'$-expansion in string theory.

Finally, with all its attractive features, this extended formalism is only intended as a macroscopic approximation of the 'real' microscopic nature of spacetime. The approach still needs indeed a quantum mechanical input in order to find the structure of each of the non-vanishing scalar functionals left unspecified inside the entropy functional. However, this feature only makes the approach even more interesting since it indicates exactly how the quantum nature of spacetime might manifest itself through the entropy functional.

\textbf{Acknowledgments:} I would like to thank Prof. Khireddine Nouicer for having encouraged me to extend the formalism developed in \cite{HammadBH,Hammad+Erratum} to include modified gravity theories. I would like to thank the anonymous referee for his/her helpful comments.

\end{document}